\documentclass[twocolumn,english,aps,pra,showpacs]{revtex4}
\usepackage[T1]{fontenc}
\usepackage[latin1]{inputenc}
\usepackage{graphicx}
\usepackage{amssymb}

\makeatletter

\input{epsf}

\usepackage{babel}
\makeatother
\begin{document}

\title{Temperature of a trapped unitary Fermi gas at finite entropy}

\author{Hui Hu, Xia-Ji Liu, and Peter D. Drummond}

\address{ARC Centre of Excellence for Quantum-Atom Optics, Department of Physics,
\\
 University of Queensland, Brisbane, QLD 4072, Australia}

\date{\today{}}

\begin{abstract}
We present theoretical predictions for the equation of state of a
harmonically-trapped Fermi gas in the unitary limit. Our calculations
compare Monte-Carlo results with the equation of state of a uniform gas using three
distinct perturbation schemes. We
show that in experiments the temperature can be usefully calibrated
by making use of the entropy, which is invariant during an adiabatic
conversion into the weakly-interacting limit of molecular BEC. We
predict the entropy dependence of the equation of state. 
\end{abstract}

\maketitle

\section{Introduction}

The past two years have witnessed many exciting developments in experiments
on dilute Fermi gases of ultracold $^6$Li and $^{40}$K atoms \cite
{formation,dsty,ens,momt,cm,rf,vortex,heat}. In these systems the
interaction strength between atoms, represented dimensionlessly as $k_Fa$,
where $k_F$ is the Fermi wave-vector and $a$ is $s$-wave scattering length,
can be varied arbitrarily via a Feshbach resonance. One can therefore access
the strongly interacting regime of $-1<1/k_Fa<1$, where a smooth crossover
from a Bardeen-Cooper-Schrieffer (BCS) superfluidity to a Bose-Einstein
condensate (BEC) occurs. Of particular interest is the unitary limit with
negligible interaction range and divergent scattering length ($k_Fa\simeq
\infty $), in which a universal behavior is expected \cite{uniTh,uniEx}.
This remarkable feature renders the unitary Fermi gas an intriguing
many-body system.

Evidence for the onset of fermionic superfluidity at the BCS-BEC crossover has been  found in several ground-breaking experiments. These have measured condensate
formation \cite{formation}, density \cite{dsty} and momentum distributions 
\cite{momt}, collective excitations \cite{cm}, RF spectroscopy \cite{rf},
and vortices \cite{vortex}. A thermodynamic measurement of the heat capacity
of $^6$Li atoms has also been performed very close to the unitary limit \cite
{heat}, showing an abrupt jump at an estimated temperature of about $0.27T_F$%
, where $T_F$ is the Fermi temperature. However, there is no
model-independent method to measure the temperature of a strongly
attractive, deeply degenerate Fermi gas. In experiments the gas is
characterized indirectly by an empirical temperature (or thermometry)
obtained by fitting integrated one-dimensional (1D) density profiles to an
ideal Thomas-Fermi (TF) distribution \cite{heat}. For these isolated
systems, the entropy is a more readily measurable than the temperature. The
goal of this work is to predict the equation of state in terms of measurable
quantities like the entropy.

In contrast to these rapid experimental advantages, theoretical progress on
the crossover is quite limited \cite
{eagles,leggett,nsr,randeria,haussmann,strinati,hld,ohashi,qmc}. In the
strongly correlated unitary limit, it is extremely difficult to
construct a quantitative theory in terms of a well-defined small parameter,
especially at finite temperature. Therefore, current theoretical
understanding of experimental data relies heavily on the simplest BCS mean
field picture \cite{hu,kinnunen,heat}. In particular, the heat capacity
measurement has been explained by a crossover theory \cite{heat}, where a
BCS-like ground state is generalized to accommodate thermal bosonic degrees
of freedom in the long wavelength limit \cite{chen}.

In the unitary limit of a Fermi gas there are strong pair fluctuations in
the {\it T}-matrix approximation {\em beyond} the mean-field level. These
must be included to predict the energy as a function of entropy in the
unitary regime. The entropy in turn is a function of more readily measured
temperatures in well-understood weakly interacting regimes, which are
accessible via adiabatic changes in the coupling constant. The combination
of measured temperatures of molecular BEC together with the state-equation,
also allows one to estimate the corresponding strongly-interacting
temperatures. While entropy invariance under adiabatic passage is
well-understood, the crucial step here is to obtain a reliable
state-equation below threshold that covers the entire range from weak to
strong interactions.

Without a controllable small parameter in the unitary regime, the
determination of the equation of state is by no means a trivial task. In
general, there are a number of alternative {\it T}-matrix schemes that can
be implemented to tackle the crossover problem, and there is no {\em a priori%
} judgement of which one is the most accurate. In this regard, it is of
particular relevance that Monte Carlo simulations have been performed in the
unitary limit at finite temperature \cite{bulgac}, which present useful
benchmarks on the validity of different approximations. Here we present a
comparative study of the equation of state of a {\em uniform} gas using
three candidate {\it T}-matrix schemes, and show that, apart from a small
region around the transition temperature, a below threshold version of
the Nozi\`{e}res and Schmitt-Rink (NSR) scheme seems to be the optimal choice
for the calculations \cite{nsr} of the type required for
thermometry. We then incorporate the effect of an harmonic trap using a
local density approximation that holds for large particle number.

We characterize the entropy and temperature of the strongly interacting
system by an isentropic sweep to the weakly interacting molecular BEC
regime, and a consequent determination \cite{dsty,rf,carr} of the entropy $S$
from the final temperature $T^{\prime }$. The $T^{\prime }$ dependence of
the equation of state is predicted. Alternatively, we can view this as a
predicted equation of state in terms of the entropy, which can be readily
compared to completely model-independent experimental observations of
quantities like the system energy as a function of entropy.

The manuscript is organized such that the equation of state of a uniform
unitary Fermi gas is first determined by comparing three different
perturbation schemes, and then used to calculate the equation of state of a
trapped Fermi gas within local density approximation. In turn the resulting
unitary entropy is used to match the entropy of a weakly interacting Bose
gas to obtain the desired isentropic thermometry. Finally, a summary and
conclusions are given.

\section{Equation of state of homogeneous unitary gases}

To describe a gas of $^{6}$Li atoms at a broad Feshbach resonance $B_{0}=834$
G, we adopt a single-channel model \cite{ho,lh}, in which the attractive
interatomic interaction is characterized by a contact potential of strength $%
U=4\pi\hbar^{2}a/m$. In the {\em normal} state above the superfluid
transition temperature $T_{c}$, the model can be treated perturbatively with
the inclusions of strong pair fluctuations in the many-body {\it T}-matrix
approximation \cite{chen}, which leads to a two-particle propagator $%
\chi\left({\bf q},i\nu_{n}\right)$ given by 
\begin{equation}
\chi\left({\bf q},i\nu_{n}\right)=\frac{1}{\beta}\sum_{{\bf k}%
,i\omega_{m}}G_{a}\left({\bf k},i\omega_{m}\right)G_{b}\left({\bf q-k}%
,i\nu_{n}-i\omega_{m}\right),  \label{2p-propagator}
\end{equation}
and a corresponding self-energy of form, 
\begin{equation}
\Sigma\left({\bf k},i\omega_{m}\right)=\frac{1}{\beta}\sum_{{\bf q},_{i\nu
n}}\frac{U}{1+U\chi\left({\bf q},i\nu_{n}\right)}G_{c}\left({\bf q-k}%
,i\nu_{n}-i\omega_{m}\right).  \label{self-energy}
\end{equation}
where $\omega_{m}\equiv\left(2m+1\right)\pi/\beta$ and $\nu_{n}\equiv2n\pi/%
\beta$ are, as usual, the fermionic and bosonic Matsubara frequencies with
inverse temperature $\beta=1/k_{B}T$. The superscript $a$, $b$, and $c$ in
the above two equations can be set to {}``0'', indicating a non-interacting
Green function $G_{0}\left({\bf k},i\omega_{m}\right)=1/[i\omega_{m}-%
\hbar^{2}{\bf k}^{2}/2m+\mu]$. If this superscript is absent, we refer to a
fully dressed (interacting) Green function, and thereby a Dyson equation, $%
G\left({\bf k},i\omega_{m}\right)=G_{0}\left({\bf k},i\omega_{m}%
\right)/[1-G_{0}\left({\bf k},i\omega_{m}\right)\Sigma\left({\bf k}%
,i\omega_{m}\right)]$, is required to self-consistently determine $G$ and $%
\Sigma$. The only free parameter is the chemical potential $\mu$, which is
fixed by using the number equation $n=2/\beta\sum_{{\bf k}%
,i\omega_{m}}G\left({\bf k},i\omega_{m}\right)$. According to the different
combination of subscripts of $a$, $b$, and $c$, on general grounds we may
expect three obvious choices of the {\it T}-matrix approximation, for which
a nomenclature of $\left(G_{a}G_{b}\right)G_{c}$ will be used. These
alternative choices distinguish themselves by the level of self-consistency
contained in the Green function.

The simplest choice, $\left( G_0G_0\right) G_0$, was proposed in a seminal
paper \cite{nsr} by NSR, although these authors approximated the Dyson
equation using a leading order series, {\it i.e.}, $G=G_0+G_0\Sigma G_0$.
This fully non-self-consistent scheme includes the least Feynman diagrams.
In the other extreme limit, one can consider a fully self-consistent {\it T}%
-matrix approximation of $\left( GG\right) G$, which is a so-called
conserving approximation \cite{haussmann}. An alternative intermediate
scheme is to use an asymmetric form for the two-particle propagator, {\it %
i.e.}, $\left( GG_0\right) G_0$ \cite{chen}. These candidate versions of 
{\it T}-matrix approximation have been extensively discussed for the
attractive Hubbard model in the context of high-$T_c$ superconductors. The
applicability of these theories to the single-channel model is under active
debate, due to the complexity of numerical calculations. In this work, we
have resolved the numerical difficulties by using a technique of adaptive
step Fourier transformations \cite{haussmann,lh}, without additional
approximation.

The situation in the {\em broken-symmetry} state is more subtle. Below $%
T_{c} $, the denominator in Eq. (\ref{self-energy}) develops a pole that
signals the emergence of superfluid phases. To remove the instability, one
may extend the different {\it T}-matrix approaches by using a BCS mean-field
ground state as the starting point for a perturbation theory. For the$%
\left(G_{0}G_{0}\right)G_{0}$ scheme, this strategy has been recently
adopted and investigated in depth by Strinati {\it et al.} \cite{strinati}
and by the present authors \cite{hld}. The advantage of the latter study is
that a modified number equation was employed to obtain an (approximately)
correct molecular scattering length, which therefore makes the theory
quantitatively reliable over the entire crossover regime \cite{hld}. The
other alternative versions of $\left(GG_{0}\right)G_{0}$ and $%
\left(GG\right)G$ along this line have not been explored yet. However, on
physical grounds, below $T_{c}$ one may expect only a slight difference
among these schemes, as the ground state is well stabilized by the
mean-field gap.

\begin{figure}
\begin{center}
\includegraphics[%
  width=8.5cm]{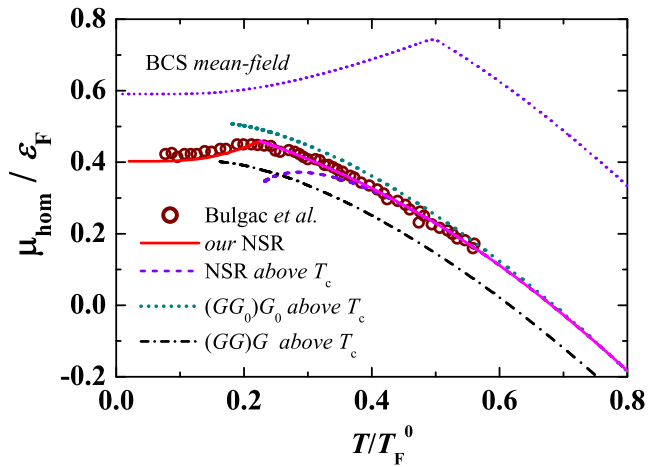}

\caption{(color online) Chemical potential of a uniform Fermi gas 
in the unitary limit as a function of temperature, 
in units of Fermi energy $\epsilon_{F}=\hbar^{2}k_{F}^{2}/2m=k_{B}T_{F}^{0}$.
Lines plotted are: our proposed results (solid lines); NSR $\left(G_{0}G_{0}\right)G_{0}$
predictions in Ref. \cite{nsr,randeria} (dashed line); asymmetric
$\left(GG_{0}\right)G_{0}$ scheme (dotted line); fully self-consistent
$\left(GG\right)G$ scheme (dot-dashed line). All these
results are contrasted to the Monte Carlo data from Ref. \cite{bulgac}
(open circles).}
\label{fig1}

\end{center}
\end{figure}

Figure (1) presents a comparative study of these candidate {\it T}-matrix
approaches for a uniform unitary gas at finite temperature. In the
superfluid state, we use the NSR-like formalism implemented in Ref. \cite
{hld}. The calculated chemical potentials of the gas are compared with a
recent Monte Carlo simulation from Ref. \cite{bulgac}. Just above the
transition temperature $T_c\sim 0.2T_F^0$, all approximations appear poor,
with errors of around $\pm 10\%$. In particular, the NSR approach suffers
from an unphysical drop with decreasing temperature, which should be due to
the lack of self-consistency in the two-particle propagator. At high
temperatures above $0.35T_F^0$, the NSR result turns out to be very
accurate. This observation, together with the excellent agreement between
the NSR prediction and Monte Carlo data below $T_c$, suggest that the fully
non-self-consistent NSR scheme is generally accurate, apart from the regime
just above $T_c$. The spurious structure about $T_c$ in this scheme might be
avoided by a phenomenological interpolation between the results at $T_c$ and
at a high temperature $\sim 0.4T_F^0$. The outcomes with this idea have been
denoted in the figure by solid lines, and referred to later as {}``our NSR''
results. From this, we are able to calculate the energy and entropy.

We emphasize that the equation of state obtained in this manner is certainly
not {\it ab initio}, but it should be quantitatively valid, provided the
Monte Carlo simulation data are accurate. More reliable Monte Carlo
simulations with larger lattices may change these conclusions slightly in
future, especially in the vicinity of $T_c$, where already there are reports
of small discrepancies between the simulations carried out with different
lattices and algorithms. The important issue here is that while there are
discrepancies {\em above} $T_c$ which make it difficult to decide which
approximation is better, the below $T_c$ region which is crucial for finite
entropy thermometry methods, shows excellent agreement between Monte-Carlo
data and our $\left( G_0G_0\right) G_0$ scheme.

\begin{table}[h]
\begin{tabular}{ccccc}
\\ \hline \hline
Methods & $T_c/T_{F}^{0}$ & $\mu(T_c)/\epsilon_F$ & $E(T_c)/(N \epsilon_F)$
& $S(T_c)/(Nk_B)$ \\ \hline
BCS-MF & 0.50 & 0.743 & 1.022 & 1.92 \\ 
$(G_0G_0)G_0$ & 0.225 & 0.340 & 0.290 & 0.64 \\ 
$(GG_0)G_0$ & 0.178 & 0.508 & N.A. & N.A. \\ 
$(GG)G$ & 0.150 & 0.404 & N.A. & N.A. \\ 
{\it our} NSR & 0.225 & 0.459 & 0.400 & 0.91 \\ 
Monte Carlo & 0.23(2) & 0.45 & 0.41 & 0.99 \\
\hline \hline
\end{tabular}
\caption{The superfluid critical temperature for a homogeneous unitary Fermi
gas obtained by various methods. The chemical potential, total energy, and
entropy at the critical temperature are also listed. We note that all the
methods, except the $(GG_0)G_0$ scheme, satisfy an exact identity $P=2/3E$
that holds in the unitary limit, where $P=-\Omega=-(E-TS-\mu N)$ is the pressure of the
gas. The energy and entropy for the $(GG_0)G_0$ and $(GG)G$ approximations
are not available.}
\end{table}

We close this section by listing in Table I the predictions for the
superfluid transition temperature that were obtained by the various models
we have discussed. The chemical potential, the total energy and the entropy
at the transition temperature are also outlined.

\section{Equation of state of trapped unitary gases}

With the knowledge of equation of state of a uniform gas, we incorporate the
effects of harmonic traps by using the local density approximation, which
amounts to determining the global chemical potential from the local
equilibrium condition, 
\begin{equation}
\mu=\mu_{\hom}\left[n({\bf r}),\frac{T}{T_{F}}\frac{T_{F}}{T_{F}^{0}\left(%
{\bf r}\right)}\right]+\frac{m\omega_{r}^{2}}{2}\left(r^{2}+\lambda^{2}z^{2}%
\right),  \label{mu}
\end{equation}
with a density $n({\bf r})$ normalized to the total number of atoms, $\int
n\left({\bf r}\right)d{\bf r}=N$. Here $\mu_{\hom}$ is our proposed NSR
chemical potential of a uniform gas in Fig. (1), and $\omega_{x}=\omega_{y}=%
\omega_{r}$ and $\omega_{z}=\lambda\omega_{r}$ are different frequencies for
three axis. In this case, the energy and temperature can be taken in units
of the (non-interacting) Fermi energy $E_{F}=\left(3N\omega_{x}\omega_{y}%
\omega_{z}\right)^{1/3}$ and $T_{F}=E_{F}/k_{B}$, respectively. Once $n({\bf %
r})$ are obtained, the energy and the entropy of the gas can then be
calculated straightforwardly, using $E=\int n\left({\bf r}\right)E_{\hom}d%
{\bf r}$ and $S=\int n\left({\bf r}\right)S_{\hom}d{\bf r}$, with our
proposed equation of state of a homogeneous gas, $E_{\hom}\left[n({\bf r}%
),T/T_{F}^{0}\left({\bf r}\right)\right]$ and $S_{\hom}\left[n({\bf r}%
),T/T_{F}^{0}\left({\bf r}\right)\right]$.

\begin{figure}
\begin{center}
\includegraphics[%
  width=8cm]{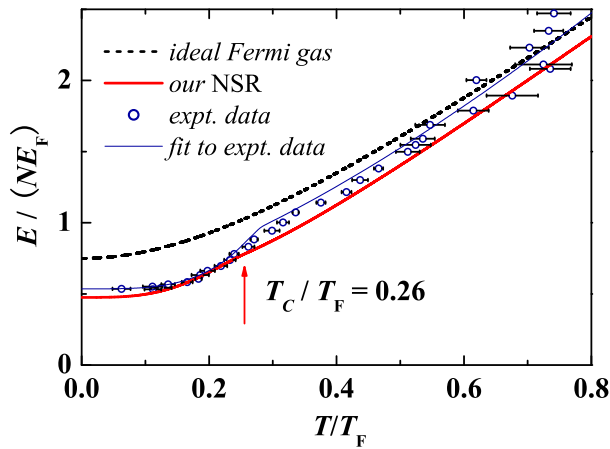}

\caption{(color online) Energy of a trapped unitary gas is plotted as a function 
of temperature (thick solid line), and compared to experimental data (open circles with
error bars). The comparison with respect to the prediction of an ideal Fermi gas (dashed
line) is also shown. The thin solid line is an empirical power law fit to the measured energies
\cite{heat}, {\it i.e.}, $E(T)=E_0[1+97.3(T/T_F)^{3.73}]$ for $ T<T_c\simeq 0.27T_F$, 
and $E(T)=E_0[1+4.98(T/T_F)^{1.43}]$ above $T_c$. Here $E_0$ is the energy at zero temperature
and $T_c=0.27T_F$ is the experimentally extracted superfluid transition temperature \cite{heat}.}

\label{fig2}

\end{center}
\end{figure}

\begin{figure}
\begin{center}
\includegraphics[%
  width=8cm]{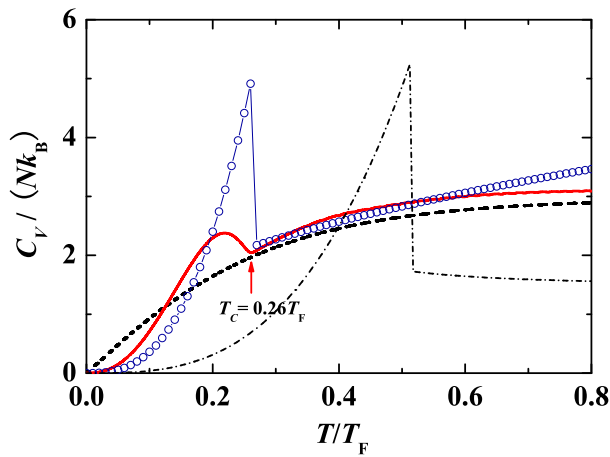}

\caption{(color online) Specific heat $C_{V}=dE/dT$ of a trapped unitary gas as 
a function of temperature. For comparison, the predictions of an ideal Fermi gas
(dashed line) and of an ideal Bose gas with number of particles $N_B=N/2$ (dot-dashed line) \cite{pethick},
together with the experimental extracted specific heat (line with open circles) are also 
presented. It should be noted that a precise determination of the specific heat from 
experimental data of energies is not permissible. The experimental specific heat shown here
is calculated from the empirical power law fit to the measured energies, which is outlined 
in Fig. 2. The resulting sharp jump at $T_c$ is thereby possibly an artifact of the fit. 
We note also that the abrupt discontinuity in the specific heat of an ideal trapped Bose 
gas at $T_c$ will be rounded off quickly with inclusions of boson-boson interactions, 
see, for exmaple, the Fig. 3 in Ref. \cite{anna}.}

\label{fig3}

\end{center}
\end{figure}

The temperature dependence of the energy and the specific heat of a trapped
unitary gas are given in figures (2) and (3), and are compared to the
experimental data obtained by Kinast {\it et al.} in the vicinity of unitary
limit ($1/k_Fa\simeq -0.03$) \cite{heat}. It is worth noting that these
experimental data are not completely raw, since the measured empirical
temperature requires a model-dependent theorectical recalibration \cite
{note1}.

Our results agree qualitatively with experimental data. In particular, our
predictions for the specific heat show an apparent superfluid transition at
about $T_c\simeq 0.26T_F$, in excellent agreement with the experimental
findings \cite{note2}. The peak structure resembles the $\lambda $ transition
for a BEC. The behavior of specific heat of a unitary Fermi gas thereby lies
between that of ideal Fermi gases and that of BECs. The value of both
theoretical and experimental $T_c$ need further refinement, as the detailed
equation of state of a uniform gas around the transition temperature is
blurred by the semi-empirical interpolation, and the experimental
temperatures are not known accurately. Another noticeable feature of our
result is that it deviates greatly from the non-interacting Fermi gas
behavior for all the temperatures considered. This is suggestive of the
strongly-correlated nature of unitary gases, and is in sharp contrast to the
prediction of mean-field crossover theories by Chen {\it et al.} \cite
{heat,chen}.

\section{Isentropic thermometry for trapped unitary gas}

A quantitative comparison between our theorectical results and experimental
findings is prohibited, as current estimates of strongly-interacting Fermi
gas temperatures are based on empirical thermometry through density profile
measurements, which are strongly dependent on the model used, but relatively
insensitive to the actual temperature \cite{note3}. A better thermometry
would be obtained by an adiabatic conversion of the unitary gas to the deep
BEC limit, and a consequent measurement of BEC temperature $T^{\prime }$ of
a resulting Bose gas \cite{carr}. Such techniques have been used in recent
experiments for temperature estimates \cite{dsty,rf}, but the conversion
between $T^{\prime }$ and the unitary temperature $T$ remains unknown. Below
we quantify this important relation using the isentropic condition $%
S^{\prime }(T^{\prime })=S(T)$.

\begin{figure}
\begin{center}
\includegraphics[%
  width=8cm]{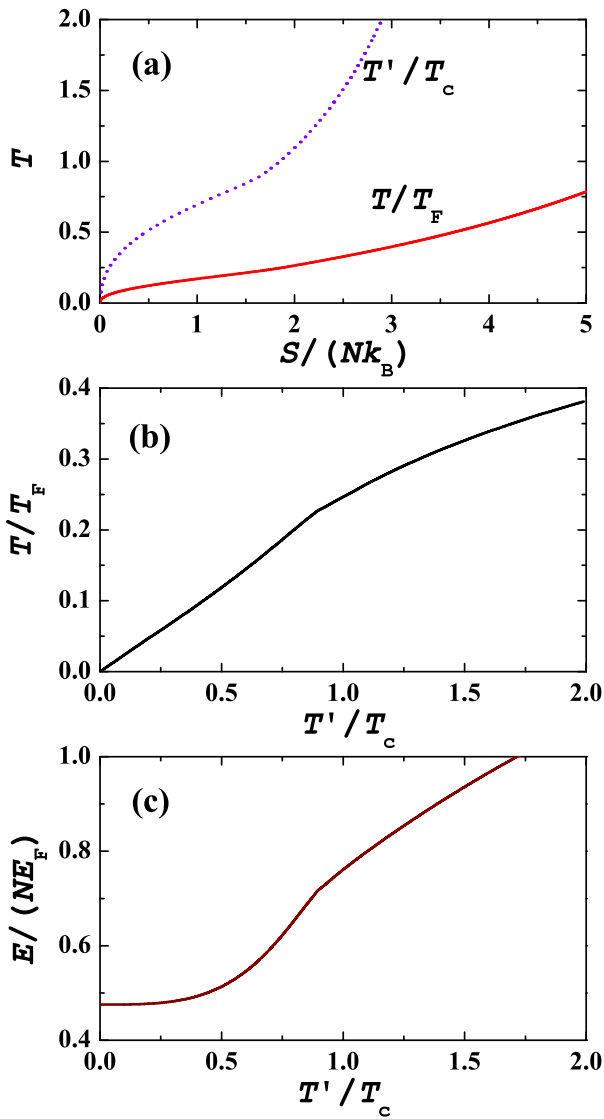}

\caption{(color online) (a) The temperature as a function of entropy in the
unitary limit (solid line) and in the deep BEC limit of $1/k_{F}a\simeq4.9$
(dashed line), with $T_{c}=0.518T_{F}$. At low temperature, our calculated
BEC entropy satisfies the scaling law for a weakly-interacting Bose
gas \cite{giorgini}: $S^{\prime}(T^{\prime})\propto$ $\left(T^{\prime}\right)^{5/2}$.
(b) Calibration curve of unitary temperature versus weakly interacting
Bose gas temperature. (c) Unitary energy as a function of BEC temperature
$T^{\prime}$. }
\label{fig4}

\end{center}
\end{figure}

To be concrete, we consider a gas of $2\times 10^5$ $^6$Li atoms with a
potential $\omega _r=1555$ Hz and $\omega _z=127$ Hz, according to the
typical setup \cite{dsty}. Assuming that the gas is swept to a field of $%
B=676$ G \cite{dsty}, we find $1/k_Fa\simeq 4.9$. Thus, the system after
sweep can be well described by a weakly-interacting Bose gas with molecular
scattering length $a_m=0.6a$. The entropy $S^{\prime }(T^{\prime })$,
calculated within the Hartree-Fock-Bogoliubov-Popov theory \cite{giorgini},
has been compared with the unitary entropy $S(T)$ in Fig. (4a). This
calibrates the unitary temperature, given in Fig. (4b). Accordingly, the
equation of state can also be determined in terms of the BEC temperature $%
T^{\prime }$, as shown in Fig. (4c). It is worth noticing that this kind of
thermometry should be very accurate, thanks to the demonstrated prefect
adiabatic sweep of magnetic fields \cite{dsty} and a precise determination
of BEC temperature \cite{gerbier}.

\section{Summary}

In the present paper we have studied the equation of state of a trapped
unitary gas with a perturbation theory that includes fluctuations
beyond the mean-field. Our investigations, in contrast to mean-field theory \cite
{eagles,leggett,chen}, are in excellent agreement with Monte Carlo
simulations below threshold, although there are still uncertainties 
of order $10\%$ around threshold. We find a qualitative agreement between our
theoretical predictions and the experimental findings. A fully quantitative
comparison is not yet possible, as the experimental temperatures are not
directly determined. We reach the conclusion that the
empirical thermometry adopted in the experiment can be improved by using
adiabatic sweeps at constant entropy to the weakly interacting molecular
superfluid regime. The equation of state based on this type of thermometry
has been predicted, allowing for experimental tests.

We emphasize that while our NSR result for the equation of state is very
accurate at both low and high temperatures, its accuracy around the critical
temperature $T_c$ requires a further improvement beyond {\it T}-matrix
approximation. In terms of functional-integral methods \cite{randeria}, our
NSR approach can be identified as a Gaussian expansion of the action of
Cooper-pairs about the saddle point (mean-field solution). Therefore, a
systemmatic refinement of equation of state around $T_c$ might be achieved
by expanding the action to fourth or higher order about the saddle point. 
We will address this issue in a later publication.

Our calculated equation of state can be used to predict other
observables of experimental interest, such as the first sound and second
sound velocities of trapped unitary Fermi gases at finite temperature \cite
{heiselberg,griffin}, which soon may be measured. These sound modes,
particularly the second sound, may provide us more accurate details about
the equation of state in the unitary limit, and therefore discriminate 
various approximations that we have marked. This extension is under
investigation and will be reported elsewhere.

\section*{Acknowledgement}

We are indebted to J. E. Thomas and J. Kinast for discussions. Funding for
this work was provided by an Australian Research Council Center of
Excellence grant.


\begin{thebibliography}{99}
\bibitem{note3}  The resulting empirical temperature itself is
model-independent. However, to recover the actual temperature, an assumption
is used that the full $3D$ density distribution in the unitary limit is
close to an ideal TF functional form. We have checked this, and found that
this essential assumption is not well-satisfied either in the mean-field or
NSR theories. Hence, the mapping from empirical to actual temperatures can
have large random and systematic errors.
\end{thebibliography}
\end{document}